# Efficient Microwave Absorption in Thin Cylindrical Targets: Experimental Evidence

A. Akhmeteli[1], N. G. Kokodiy[2], B. V. Safronov[2], V. P. Balkashin[2], I. A. Priz[2], A. Tarasevitch[3]

[1]LTASolid Incorporated, 10616 Meadowglen Ln 2708, Houston, TX 77042, USA
E-mail: *akhmeteli@ltasolid.com*

[2]Kharkov National University, Ukraine, Kharkov, Svobody sqw., 4
E-mail: *kokodiy.n.g@gmail.com*

[3]University of Duisburg-Essen, Institute of Experimental Physics, Duisburg, Germany
Lotharstr. 1, 47048 Duisburg, Germany
E-mail: *alexander.tarasevitch@uni-duisburg-essen.de*

**Abstract:** Significant (4%) absorption of microwave power focused on a thin fiber (the diameter is three orders of magnitude less than the wavelength) is demonstrated experimentally.

## Introduction

This work is based on some relatively new results for the old problem of diffraction of electromagnetic radiation on a circular cylinder.

Shevchenko [1] derived optimal conditions of absorption of a plane electromagnetic wave in a thin conducting wire. His results were experimentally confirmed by Kuz'michev, Kokodii, Safronov, and Balkashin [2]. Independently of work [1], Akhmeteli demonstrated theoretically that a significant part (tens percent) of the power of a cylindrical electromagnetic wave converging on a thin conducting wire can be absorbed in the wire under conditions close to those derived by Shevchenko (see [3-5] and references there). Target irradiation geometry of work [3-5] allows efficient non-resonant heating of thin wires by broad electromagnetic beams. In this article we provide experimental confirmation of this result.

## Experiment

The geometry of the experiment is shown in Fig. 1.

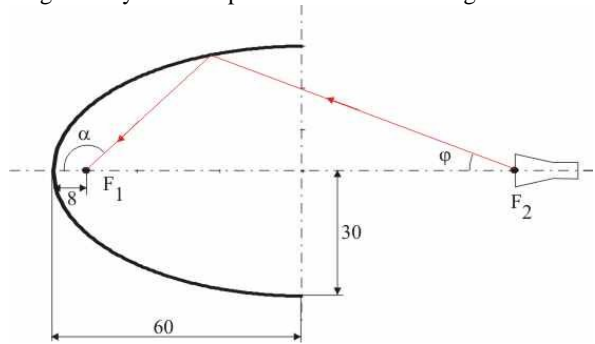

Fig. 1. The geometry of the experiment

The reflector is a copper ellipsoid of revolution, its major semi-axis is 60 mm, its minor semi-axis is 30 mm. Focus $F_1$ is at a distance of 8 mm from the vertex. The wire (fiber) is placed in this focus. The 32x22 mm aperture of the feed horn is at the other focus ($F_2$) (this is not the optimal location). The long side of the aperture is horizontal. The wavelength is 7.7 mm (39 GHz). The waveguide dimensions are 7.2 x 3.4 mm, the horn length is 130 mm. The radiation is polarized vertically (the electric field is vertical). The radiation power is 55 mW.

The generator part of the experimental set is shown in Fig. 2. The power was measured with a thermistor power meter included in the waveguide system. A ferrite isolator was used to suppress the reflected wave in the waveguide.

The resistance of the wire was measured with a high accuracy digital ohmmeter.

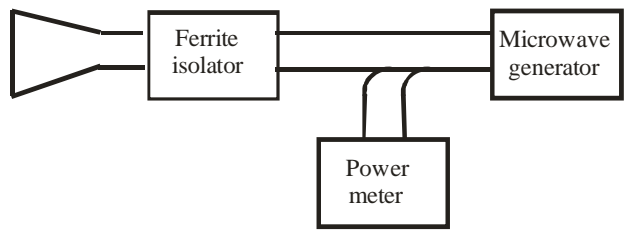

Fig. 2. The scheme of generator part

In our experiments, the power $P$ absorbed by the wire was determined by measuring the change $\Delta R$ of the wire (fiber) resistance. The average wire temperature increase $\Delta T$ corresponding to $\Delta R$ was calculated as

$$\Delta T = \Delta R/(\alpha_r R_0).$$

Based on this temperature increase, the power absorbed in the wire was determined as

$$P = \alpha_p L \Delta T.$$

The above value $\alpha_r$ is the temperature coefficient of resistance, $R_0$ is the initial resistance of the wire, and $\alpha_p$ is the linear heat exchange coefficient.

The experiments were carried out with a platinum wire and a carbon fiber. The platinum wire had the diameter $D = 3.5$ µm, length $L = 15$ mm, and resistance $R_0 = 138.35$ Ohm.

When the wire is parallel to the electric field in the horn (vertical), its resistance increases by $\Delta R = 1.19$ Ohm, which corresponds to average heating of
$$\Delta T = 2.15° \text{ C}$$
and
$$P \approx 1 \text{ mW}.$$
The following values were used for the platinum wire: $\alpha_r = 0.004$ deg$^{-1}$ and $\alpha_p = 0.03$ W/(m·deg).

For the orthogonal wire orientation (horizontal), much smaller power $P \approx 0.2$ mW is absorbed ($\Delta R = 0.23$ Ohm, $\Delta T = 0.42°$C)

Without the reflector and with the vertical wire orientation, only about 17 µW is absorbed ($\Delta R = 0.02$ Ohm, $\Delta T = 0.04°$ C).

Similar results were obtained with the carbon fiber [6], diameter 11 µm, length 15 mm, initial resistance 2816 Ohm.

When the fiber is parallel to the electric field in the horn (vertical) and located in $F_1$, the resistance change (reduction) was measured to be $\Delta R = -3.0$ Ohm. According to our measurements, $\alpha_r = -2.11 \cdot 10^{-4}$ deg$^{-1}$. This corresponds to $\Delta T = 5.05°$ and the absorbed power of 2.3 mW (approximately 4% of the total feed power).

The drop of the absorbed power for the horizontal fiber orientation was much more pronounced compared to the case of the platinum wire. The corresponding change of resistance $\Delta R$ could not even be measured.

With the reflector removed and vertical fiber, the resistance changes by $\Delta R = -0.3$ Ohm, and the absorbed power is 230 µW.

## Conclusion

The experimental results were compared with the results of computations. The feed horn fields (incident on the reflector) were computed using the formulae for the far zone of a pyramidal horn [7]. The reflected fields for the ellipsoid were estimated using methods of physical optics [8], although these methods are not quite adequate for the problem, as the radii of curvature of the reflector are not much greater than the wavelength everywhere. It is assumed that the linear absorbed power in any section of the wire (fiber) is the same as for an incident cylindrical wave with the same electric field component along the axis on the axis of the wire (fiber). For vertical platinum wire and carbon fiber, the computed absorbed power is 0.6 mW and 5.1 mW, respectively. These values differ from the experimental ones (1 mW and 2.3 mW, respectively) approximately by a factor of 2, but the computed value is less than the experimental one for platinum wire, whereas the computed value is greater than the experimental one for the carbon fiber. Overall, the computed values are in good agreement with the experimental ones. The discrepancy may be caused by the limitations of the physical optics approximation. . Indeed, strictly speaking, these methods are not quite adequate for the problem, as the radii of curvature of the reflector are not much greater than the wavelength everywhere. The calculation also does not take secondary reflections and the effect of the wire holders into account. Some of these limitations were removed using a much larger reflector. The results of these experiments, where up to 10% of the feed power and more was absorbed in the carbon fiber, will be presented later.